# The human-AI relationship in decision-making:

AI explanation to support people on justifying their decisions

Juliana Jansen Ferreira

IBM Research Brazil, jjansen@br.ibm.com

Mateus Monteiro

IBM Research Brazil, Federal Fluminense University, msmonteiro@ibm.com

The explanation dimension of Artificial Intelligence (AI) based system has been a hot topic for the past years. Different communities have raised concerns about the increasing presence of AI in people's everyday tasks and how it can affect people's lives. There is a lot of research addressing the interpretability and transparency concepts of explainable AI (XAI), which are usually related to algorithms and Machine Learning (ML) models. But in decision-making scenarios, people need more awareness of how AI works and its outcomes to build a relationship with that system. Decision-makers usually need to justify their decision to others in different domains. If that decision is somehow based on or influenced by an AI-system outcome, the explanation about how the AI reached that result is key to building trust between AI and humans in decision-making scenarios. In this position paper, we discuss the role of XAI in decision-making scenarios, our vision of Decision-Making with AI-system in the loop, and explore one case from the literature about how XAI can impact people justifying their decisions, considering the importance of building the human-AI relationship for those scenarios.

Human-centered computing • Human-computer interaction (HCI)• HCI theory, concepts and models

**Additional Keywords and Phrases:** eXplainable AI, AI design, decision-making, high-stakes decision-making, human-AI relationship

**ACM Reference Format:**
Juliana Jansen Ferreira, and Mateus Monteiro. 2021. The human-AI relationship in decision-making: AI explanation to support people on justifying their decisions. In Transparency and Explanations in Smart Systems (TExSS) workshop, held in conjunction with ACM Intelligent User Interfaces (IUI), April 13 – 17, 2021, virtual event. 9 pages.

## 1 INTRODUCTION

The use of Artificial Intelligence (AI) is growing exponentially, and with that, the challenges and opportunities of how it can impact people's lives [5][14][15]. The explainability dimension of AI, eXplainable AI (XAI), is becoming an indispensable and central feature of any AI-System. Considering that, the human in the loop in this human-AI partnership cannot be left out of the context to advance research about the impacts of AI on real-world problems [5][12][24]. While Machine Learning (ML) techniques and methods are resourcefully dealing with many data, humans' input adds meaning and purpose to that data [5][14]. XAI plays a vital role in the human reasoning process about AI's results since complex ML models can be incomprehensible for those who interact with AI's outcomes [2]. There is a fundamental challenge in balancing AI's powerful capabilities with designing technology that empowers people. Besides, people should be able to understand how technology may affect them, trust it, and feel in control [5]. The best results come from the partnership between AI and people, enabling new ways for the human brain to think and computers to process data [14].

Previous work observed in the literature an occurrence of the terms "transparency" and "interpretability" as synonyms of XAI and usually related to algorithms or ML models [14]. One example of transparency as an XAI synonym is shown in [28], presenting research on using transparency to explore intelligibility through interaction and instructions for blind users, resulting in positive indications for a more useful and effective computer vision system. For interpretability, the authors in [11] explore visualization and model interpretability to promote model verification and debugging methods using a visual analytics system. Considering the human-AI partnership as a critical aspect of XAI success, interpretability and transparency are just two of many characteristics that can be considered for an XAI approach [1][3][19]. They might be the relevant characteristics for a specific type of people - interpretability is often associated with data scientists as users [14] - or a particular goal – transparency is usually related to legal aspects of XAI [18]. Still, they are not enough for any XAI approach.

In the context of decision-making, particularly for high-stakes processes (i.e., decision-making processes that may have a significant impact on people's lives) [8], the human-AI partnership is even more crucial for the process' outcome. In high-stake decision-making, like medical diagnosis and criminal investigations, AI can provide a different way to assess and classify a large amount of diverse data [25][32], but this is only a part of the decision-making process. In the end, the doctor or the criminal investigator needs to decide about a person's heath or testify in a court of law about someone else's future. In those contexts, decision-makers need to justify their decision to others. If that decision is somehow based on or influenced by an AI-system outcome, the explanation about how the AI reached that outcome is key to building trust between humans and AI. Those decision-makers must be conscious of the role that the AI-system might have in their final decision.

In this position paper, we discuss the role of XAI in decision-making scenarios, our vision of *Decision-Making with AI-system in the loop* cycle and explore one case presented in the literature about how XAI can impact people justifying their decisions, considering the importance of building the human-AI relationship for those scenarios. We selected one interesting case from the high-stakes medical domain decision-making [32] to explore and discuss the impact that XAI might have in providing inputs for people to justify their decisions, considering the importance of building the human-AI relationship. In the following sections, we present our views and references about decision-making and AI. Then, we present and discuss the aspects of XAI in the case of a high-stakes decision-making scenario. And finally, we conclude the paper with our final remarks about the XAI in decision-making and the human-AI relationship.

## 2  DECISION-MAKING AND AI

As Artificial Intelligence (AI) technology gets more intertwined with every system, people use AI to make decisions on their everyday activities. AI having a role in people's decisions in contexts like a Netflix recommendation; the impacts are not very significant. The user may not be satisfied with the application's user experience, but it will not harm that user or any other person. But for high-stakes decision-making processes [6], like medical and judicial scenarios, the role AI plays must be clear for decision-makers so they can reach a decision considering all the impact factors.

People make decisions and, usually, they need to explain their decision to others or in some matter. It is particularly critical in contexts where human expertise is central to decision-making. To justify their decisions with AI-System support, people need to understand how AI is part of that decision. Considering AI in supporting business decision-making in general, companies that want to stand out in their industries must make the most out of the combination of domain data and domain expertise [20][31]. For high-stakes decision-making processes, the human and social impacts of the decision-making process must be central for any XAI approach designed for AI-Systems to support those processes [8][25][32].

For high-stakes decision-making processes or other decision-making processes with crucial business impacts, decision-makers will still keep the responsibility of having the final word. But with the use of AI-Systems as an empowerment tool for people in those scenarios [5], AI output can be part of or influence the final decision. Therefore, decision-makers need to understand how the AI-system got to its outcomes. In this scenario, the demand for XAI rises and becomes imperative for decisions people make in different processes [5][29]. Many important decisions are now made through an "algorithm-in-the-loop" process where machine learning models inform people. Therefore, there is an urgent need to expand our analyses of algorithmic decision-making aids beyond evaluating the models themselves to investigating the full socio-technical contexts in which people and algorithms interact [6].

### 2.1 AI in the decision-making cycle

We have been investigating the role of an AI-system in a decision-making process that relies heavily on experts' knowledge about the domain [4][14][15][16][23][30]. In this case, the AI-system provides outputs for a part of the process, and, in the end, the decision-maker needs to justify his/her decision to other people (i.e., their expert peers, their superiors, etc.). As a business decision-making scenario, not a high-stakes process, it provided us with knowledge and insights about the human-AI relationship as a whole. Besides, it informed us about the design and development of an AI-system for supporting decision-making processes.

In the context of that research, we have a broader view of ***Decision-Making with AI-system in the loop* cycle** (Figure 1). We start considering the role of designers and developers of AI-Systems in that cycle (Figure 1-A). There are more than just ML and data in an AI-system design. Therefore, all the parties involved in the development of that system are responsible for assessing and evaluating how that piece of technology can impact the decision-making process. Here, we can address questions like *"How AI Developers identify unfairness and bias in the data?"* and *"How an AI-systems user experience might impact decision-makers to reach decisions?"*. We have some findings and insights about this part of the cycle in our previous research [4][14][23][30].

Another player in the cycle is the decision-maker. He/She needs to establish a relationship with the AI-System to reach his/her decision (Figure 1-B). To do that, the decision-maker needs to be aware of the AI-System's role in his/her decision. Here, we place some research questions: *"How can decision-makers be aware be aware of how the AI-System impacts his/her decision?"*, *"How can decision-makers be aware of how their decision may affect others?"* and *"Can XAI provide part of the awareness? Is explaining enough to provide awareness?"*. We have some findings and insights about this part of the cycle in our previous research [16][30].

Finally, the last player we consider in our view, which might be the hardest one to consider in the cycle, is Society. It represents any person, entity, class, or group that can be affected, directly or indirectly, by the decision-making outcomes. It is strongly associated with the decision-making process domain. It might be the most demanding player to investigate in the *Decision-Making with AI-system in the loop* cycle. For this player, we pointed out some research questions: *"Who are the people affected directly or indirectly by the decision?", "What are the main factors in the decision?", "Would changing a certain factor have changed the decision?"* and *"Why did two similar-looking cases get different decisions?"*. We do not have research on this part of the cycle yet. It is part of our future research plans.

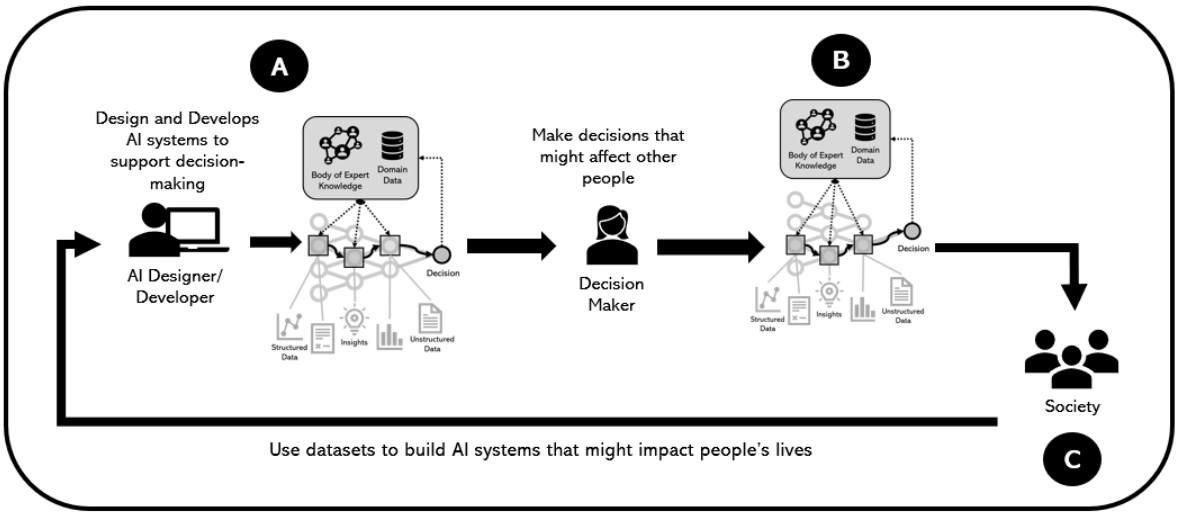

Figure 1.Decision-Making with AI-system in the loop cycle

We consider the decision-maker as the central player in the *Decision-Making with AI-system in the loop* cycle. This player is the one who knows the decision-making process and domain. He/she has a broader view of how the decisions might have an impact on real-world situations. This player should be the main informer and ally for designers and developers to build the AI-System that supports decision-making as a collaborative process where AI can help people make better decisions. Moreover, the decision-maker can provide feedback to adjust the AI-System and avoid undesired impacts, direct or indirect, on Society.

**2.2 XAI to help people justifying their decisions**

For decision-making processes in high-stakes scenarios, the justification of decisions presents as a domain practice, regardless of the technology supporting the process. The decision-maker is responsible for the final decision, but he/she needs to justify it to others [6][20]. This justification step in decision-making is part of the process. With or without any system's support, if the decision impacts a corporation or other people, it needs to be justified by the decision-maker. If that decision is somehow based on or influenced by an AI-system outcome, the explanation about how the AI reached that outcome might be part of that justification.

The definition and usage of "justification" in XAI literature are diverse. Biran and McKeown [22] describe justification as a synonym for explanation. For Hepenstal et al. [25], justification is a sub-theme of explanation, and it is defined as a piece of information that allows the justification of selected system processes. For example, in an analysis context, the user wants system understanding to support the investigation's continuation, verifying if the processes are correct. Similarly, for Adadi and Berrada [3], XAI systems must provide the required information to justify results, allowing an auditable and provable way to defend algorithmic decisions. Xie et al. [32] consider explanation and justification as two distinct categories. According to the authors, explanations enable the user to understand AI with intrinsic processes (e.g., processing pixels of an image). Justifications are extrinsic sources of information to justify AI's results (e.g., the prevalence of a diagnosis, contrastive examples from other images, patient history, among others).

There are two high-stakes decision-making scenarios where justification is an essential part of the process: medical and law domains. There are publications about clinical decision-making [7][8][17] pairing AI-Systems with doctors that make decisions about patients diagnoses and treatments. And for decision-making in law, there is work about the decision-making with AI acting as an advisor for human decision [21] and as a reliable source to support human decisions [33]. The justification is related to the role AI has in the decision, but it is still considered part of the bigger discussion about XAI. We believe decision-makers' need for justification should be a central theme for XAI on building the human-AI relationship for decision-making. Therefore, we understand that it is important to discuss XAI cases and the human-AI relationship for decision-making justification.

## 3 CASE OF XAI IN DECISION-MAKING PROCESS

To discuss the human-AI relationship for decision-making justification, we selected one case of XAI proposals from the literature to explore how XAI can impact people justifying their decisions, considering the importance of building the human-AI relationship for those scenarios. The selected case is from Xie et al. [32], from the medical domain, a high-stakes decision-making process.

In the work of Xie et al. [32], the authors present CheXplain, a system that enables physicians to explore and understand AI-enables chest X-ray analysis. The users, physicians, are a non-radiologist physician who needs to send imaging data (e.g., CXR images) of a patient to a radiologist for more information or treatment. Asking for explanations of a specific question is one of the most frequent scenarios when referring physicians contact radiologists. The goal here is not to substitute the radiologist but to augment the physician's ability to read a CXR and combine it with other exams to reach a patient's diagnosis.

Often what physicians look for is not necessarily an explanation but justification—information extrinsic to a CXR image [32]. With this proposal for an explainable AI-enabled chest X-ray analysis by interaction, physicians can now know where the AI looked and what it believes the CXR indicates to figure out the result and know exactly what is shown in CXR images, where, and why. Physicians often have a specific question for a CXR image, sometimes with important contextual information gathered from other exams (e.g., clinical exams, patient observation, history, etc.) [32]. The CXR is not the only source of information for the physician. There is contextual information, which is essential in the physician's decision-making process. The CXR is one piece of data and just a part of the decision-making process of reaching a patient's diagnosis.

We highlighted three features of CheXplain, indicated in Figure 2. The first one is mapping the CXR image with markers that point where the AI looked at and its indication of what that portion of the image is, according to its assessment (Figure 2-A). For physicians who sometimes do not know how to start looking at a CXR, this is a good starting point to combine with other information about the patient's health. Another interesting feature is the categorization of CXR's markers as *normal*, *abnormal*, and *question input related* (Figure 2-AB). It offers the physician the opportunity to focus on his/her query about the CXR and provides the CXR's overall look, as a radiologist would do. The questions physicians input into the system, as they ask radiologists, point to a hypothesis they already have about the patient condition that the CXR may help confirm or not. If the answer confirms the hypothesis, the CXR results can be used as part of the physician's justification about the next steps for the patient's treatment.

The third feature we highlighted in CheXplain is contextualizing observations with contrastive examples (Figure 2-C). This feature aligns with the practice (Physicians mentioned that comparing abnormal and normal CXR images is common in teaching radiologists). For the XAI approach, it helps align the mental model [8] for the human-AI relationship. By showing the "unlikely-definitely" range in the interface, the AI informs physicians of its references for each mark in the CXR. The physician can check if his/her own references for "unlikely" and "definitely" are aligned with the AI's. According to the authors, one way to justify the AI's result is presenting to physicians two radiographically (dis)similar images, thus enabling them to gain insight from such justifications.

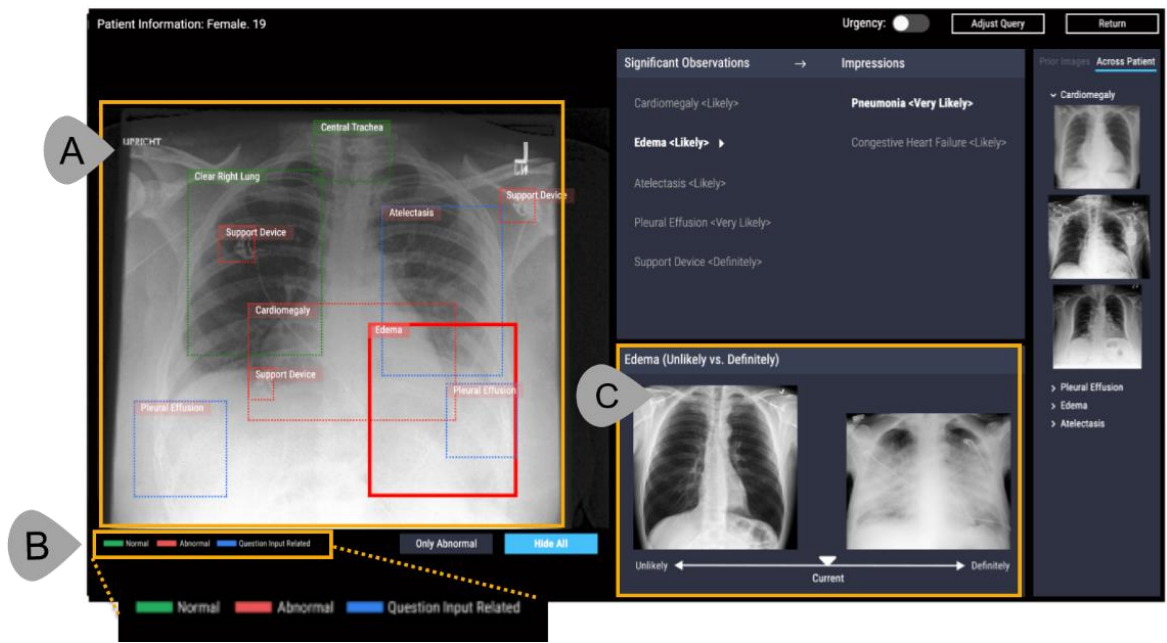

Figure 2. CheXplain: The high-fidelity prototype [32]

The CXR is part of the physician's decision-making process. The explanations provided by CheXplain provide him/her with the information to justify the decisions about a patient under examination. In this case, the highlighted features inform the physician about which parts of the image the AI points as relevant considering the physician's query. Moreover, AI presents its references with contrastive examples that offer the physician the opportunity to access the CXR results and decide if it can be used as input for justifying a patient's diagnosis or not.

## 4 FINAL REMARKS

The different AI roles for decision-making have been the subject of research for a long time (e.g., Assistant, critic, second opinion, expert) [31]. As an empowerment tool, AI can help people reach better decisions and boost their analytic and decision-making abilities. But there are still challenges when AI acts as a collaborator in the decision-making process. People struggle to understand some core elements of AI, such as ML models and algorithms [6] commonly used in explanation. Still, it is not clear if understanding those elements is always necessary to build the human-AI relationship for decision-making. An XAI approach to high-stakes decision-making processes should provide awareness about AI-system outcomes to stimulate and support the relationship building between those systems and decision-makers.

Transparency and interpretability are characteristics of XAI. They might be the relevant characteristics for a particular type of people - interpretability is often associated with data scientists as users [14] - or a specific goal – transparency is usually related to legal aspects of XAI [18]. Still, they are not enough for all XAI approach. Different ways of presenting models and structuring human-algorithm interactions may affect the quality and type of decisions made [6]. Still, it might not be the best approach for all decision-makers. The explanation should be aligned with the decision-making practice and the decision-maker skills and mental model to effectively support people in making and justifying their decisions.

The case we discussed in this paper is one example of AI explanation needs and proposals for supporting high-stakes decision-making processes. We discussed the need for AI explanation so people can justify their decisions. In this kind of process, the decision-maker has the final word in the decision, and he/she can be held accountable for that decision. Therefore, the role AI plays in that decision needs to be clear for the decision-maker. The case offers promising and validated proposals for an AI-System that participates in the decision-making process. The way AI explanation is presented can have an impact on how people justify their decisions. If the decision-maker does not trust the AI output, he/she may not use its outcomes to make the decision. The trust in AI outcomes is central for building the human-AI relationship for those scenarios. If people can understand how AI got to that outcome, maybe it will be considered part of the decision. One of the principal reasons to produce an explanation is to gain the trust of users. Trust is the primary way to enhance the confidence of users with a system [18].

As one of the authors stated in previous work [14], explanations are social, and they are a transfer of knowledge, presented as part of a conversation or interaction. They are thus shown relative to the explainer's (explanation producer) beliefs about the 'explainee's' (explanation consumer) beliefs [29]. Explanation producers are designers, developers, or other people involved in building the AI-System, and the explanation consumers are this system's users. XAI needs social mediation from technology builders to technology users and their practice [23]. We believe the explanation cannot be generic. The design of a "good" explanation needs to consider who is receiving the explanation, what for, and in which context the explanation was requested.

Society lacks both clear normative principles regarding how people should collaborate with algorithms as well as robust empirical evidence about how people do collaborate with algorithms [6]. Datasets used to train and test machine learning models reflect our Society's values and beliefs. This Society has many problems that will not vanish just because we have a "fair and unbiased" AI technology. AI-systems might be a powerful tool to expose Society's bias, unfairness, and other problems and offer people the opportunity to correct or address current issues and avoid new ones. AI systems will keep reflecting societal values and beliefs, and we should be aware and alert to mitigate any problems that might affect any person or group of people. Along with laws and regulations [18], AI-systems have the potential to support our Society to guarantee everyone's' rights and duties.

Considering that the AI collaborates with people in the decision-making process, the human-AI relationship needs a different approach than a human-human collaboration. When people need to collaborate on some endeavor, their relationship starts with a well-known mental model about the other: they are human beings. Both are somehow aware of a human being limitations and talents. Still, the particularities about that human being they are starting to have a relationship with, will be revealed as the collaboration evolves. The foundational of social psychology theories describe how collaborators form and leverage mental models of each other's capabilities [10][27]. Developing an appropriate mental model about the counterpart in collaboration is decisive for a successful relationship. People need to be aware of AI-system's capabilities to build a mental model about that system. An initial human-AI onboarding process [8] can be a way to build an initial impression and the development of appropriate mental models and strategies of use for a human-AI relationship.

In this paper, we took the perspective of the decision-maker to discuss the role of XAI, but as we discussed with our vision of *Decision-Making with AI-system in the loop* cycle, there are more people involved in that cycle. In future work, we intend to consider the perspectives of designers/developers of AI-systems and their role in the decision-making process. All people in the decision-making cycle must be aware of those that may be impacted by their decisions, keeping in mind that those people might not be so close as a physician's patient or the patient. The AI impact in Society must always be on the top priorities and concerns of those that influence the decision-making process.